\documentclass{article}
\usepackage[utf8]{inputenc}
\usepackage{graphicx}
\usepackage{caption}
\usepackage{xcolor}
\usepackage{subcaption}
\usepackage{authblk}
\usepackage{float}
\graphicspath{ {./images/} }

\title{Low field transport calculations in 2-dimensional electron gas in $\mathrm{\beta\mbox{-}(Al_{x}Ga_{1-x})_{2}O_{3}/Ga_{2}O_{3}}$  heterostructures}
\author[1]{Avinash Kumar}
\author[2]{Uttam Singisetti}
\affil[1,2]{Department of Electrical Engineering, University at Buffalo, Buffalo, New York 14260, USA}
\date{January 2020}

\begin{document}

\maketitle

\begin{abstract}
$\mathrm{\beta}$-Gallium oxide ($\mathrm{\beta\mbox{-}Ga_{2}O_{3}}$) is an emerging widebandgap semiconductor for potential application in power and RF electronics applications. Initial theoretical calculation on a 2-dimensional electron gas (2DEG) in $\mathrm{\beta\mbox{-}(Al_{x}Ga_{1-x})_{2}O_{3}/Ga_{2}O_{3}}$ heterostructures show the promise for high speed transistors. However, the experimental results do not get close to the predicted mobility values. In this work, We perform more comprehensive calculations to study the low field 2DEG transport properties in the $\mathrm{\beta\mbox{-}(Al_{x}Ga_{1-x})_{2}O_{3}/Ga_{2}O_{3}}$ heterostructure. A self-consistent Poisson-Schrodinger simulation of heterostructure is used to obtain the subband energies and wavefunctions. The electronic structure, assuming confinement in a particular direction, and the phonon dispersion is calculated based on first principle methods under DFT and DFPT framework. Phonon confinement is not considered for the sake of simplicity. The different scattering mechanisms that are included in the calculation are phonon (polar and non-polar), remote impurity, alloy and interface-roughness. We include the full dynamic screening polar optical phonon screening. We report the temperature dependent low-field electron mobility.
\end{abstract}

\section{Introduction and Overview }

$\mathrm{\beta}$-Gallium oxide ($\mathrm{\beta\mbox{-}Ga_{2}O_{3}}$) is the subject of intense research activities in recent years due to its potential applications in high voltage power devices and also high power radio frequency (RF) switching. It stands out from the list of other potential candidates (diamond, BN,AlN) because of a mature growth technology of its bulk crystals and epitaxial thin films.The power device application is due to the large Baliga's Figure of Merit (BFoM= $\mathrm{\epsilon\mu E^2}$) enabled by the large critical field strength. Experimental devices with kV breakdown voltages \cite{RN26,RN27,RN28} have been demonstrated which show the potential of the material. In addition, the calculated saturation electron velocity in ($\mathrm{Ga_{2}O_{3}}$) \cite{RN3,RN29} gives a high Johnston's Figure of Merit (JFoM= $\mathrm{Ev_{s}/2\pi}$) showing its potential for RF applications. Both MOSFETs and MESFETs RF devices with high current gain cut off frequency have been reported recently. However, in order to achieve the potential predicted by the high JFoM, it is necessary to reach the saturation velocity in the device. This in turn depends on the low field electron mobility in the device. 

The majority carriers in ($\mathrm{Ga_{2}O_{3}}$) are electrons due to lack of good p-type dopants. A low electron mobility ($\sim$200 $\mathrm{cm^{2}V^{-1}s^{-1}}$) \cite{RN16,RN37}, as compared to its other counterparts, is reported at room temperature in bulk $\mathrm{\beta\mbox{-}Ga_{2}O_{3}}$. Theoretical investigations show that the low field electron mobility is limited by the multiple polar optical phonon (POP) modes present in the material arising from the 10-atom basis \cite{RN5}. Several approaches to increase the low field mobility have been explored theoretically. An \textit{ab-initio} study on a stressed unit cell (to suppress the POP modes) of $\mathrm{\beta}$-Gallium oxide doesn't lead to any appreciable improvement in the mobility \cite{RN21}. At very high carrier concentration, the POP limited mobility was shown to increase  due to the dynamic screening of Longitudinal optical (LO) phonon modes, however the ionized impurity scattering decreases the total mobility negating any improvements due to screening. We have previously reported that the use of two-dimensional electron gas in $\mathrm{\beta\mbox{-}(Al_{x}Ga_{1-x})_{2}O_{3}/Ga_{2}O_{3}}$  heterostructures with remote doping is another way to achieve an enhancement in electron mobility. It has been predicted, taking POP and remote impurity (RI) scatterings into consideration, that the low field electron mobility in the 2DEG of such heterostructures is higher than in bulk at high 2DEG densities. There have been numerous experimental reports of 2DEG and 2DEG mobility in literature. The measured 2DEG mobilities although show an improvement over bulk mobility, do not reach the high calculated numbers in \cite{RN1}. The lower mobilities have been attributed to POP scattering and background impurity (or residual impurity) scattering. A comprehensive 2DEG mobility calculation is necessary in order to fully understand the true limits of the mobility and provide guidance to the growth of high 2DEG mobility heterostructures. 

In this work, we perform a more rigorous estimation of 2DEG mobility using \textit{ab-initio} calculated electron-phonon scattering rates. We incorporate both polar and non-polar scattering rates. In addition, we take into account alloy scattering, interface roughness (IFR) scattering which play an important role in 2DEG mobilities. A temperature dependent mobility is calculated to provide insight on the mobility limiting mechanisms. First we use the \textit{first principles} calculations of the band structure, the phonon modes and electron-phonon interaction elements. Next, we self-consistently solve the Poisson's and the Schrodinger's equations to get the confined 2DEG wavefunctions and the band profile in the 2DEG. This is described in section \ref{methods}. We take into account the dynamic screening of LO phonon modes in the calculation of POP scattering strength while for all other scattering mechanisms we assume a temperature dependent static screening. The calculation and discussion on different scattering rates is the subject of section \ref{scattering}. Here, we also discuss the anisotropy in the POP scattering due to confinement in different directions. Section \ref{mobility} describes the iterative method used to solve the Boltzmann transport equation, taking into account the elastic and inelastic scatterings, to calculate the low field mobility. In Section \ref{result}, We present the temperature dependence of low field mobility as a function of different critical device parameters and discuss the main causes of any degradation in the mobility. Finally, Section \ref{summary} concludes our work.

\section{Computational Methods}\label{methods}

\subsection{First Principle calculations}
The electronic band structure and hence the electron mass are calculated from the first principle following density function theory (DFT) framework \cite{RN5}. The calculation is performed on a three dimensional 50$\times$50$\times$50 uniform k-mesh. The energy eigen values are then interpolated using maximally localized wannier functions in order to determine the isotropic electron effective mass of 0.3m$_{e}$. A two dimensional 50$\times$50 uniform k-mesh is then used for the scattering rate calculations. The electrons are assumed to occupy the first conduction band at low electric field and hence the energies are calculated assuming effective mass approximation (spherical bands). In order to find out the lattice response of the material, we carry out the DFPT (density functional perturbation theory) calculations. The phonon dispersion is obtained on a 40$\times$40$\times$40 q-mesh. In our calculations we have not considered phonon confinement for the sake of simplicity. The long range electron-phonon matrix elements are calculated on $40\times40\times40$ uniform mesh covering 0.2 times the length of the reciprocal vectors while the short range elements are calculated on a $40\times40\times40$ uniform mesh covering 0.6 times the length of the reciprocal vectors following the procedure described in \cite{RN5}.   

\subsection{The sub-band structure}
\begin{figure}[H]
\includegraphics[width=10cm]{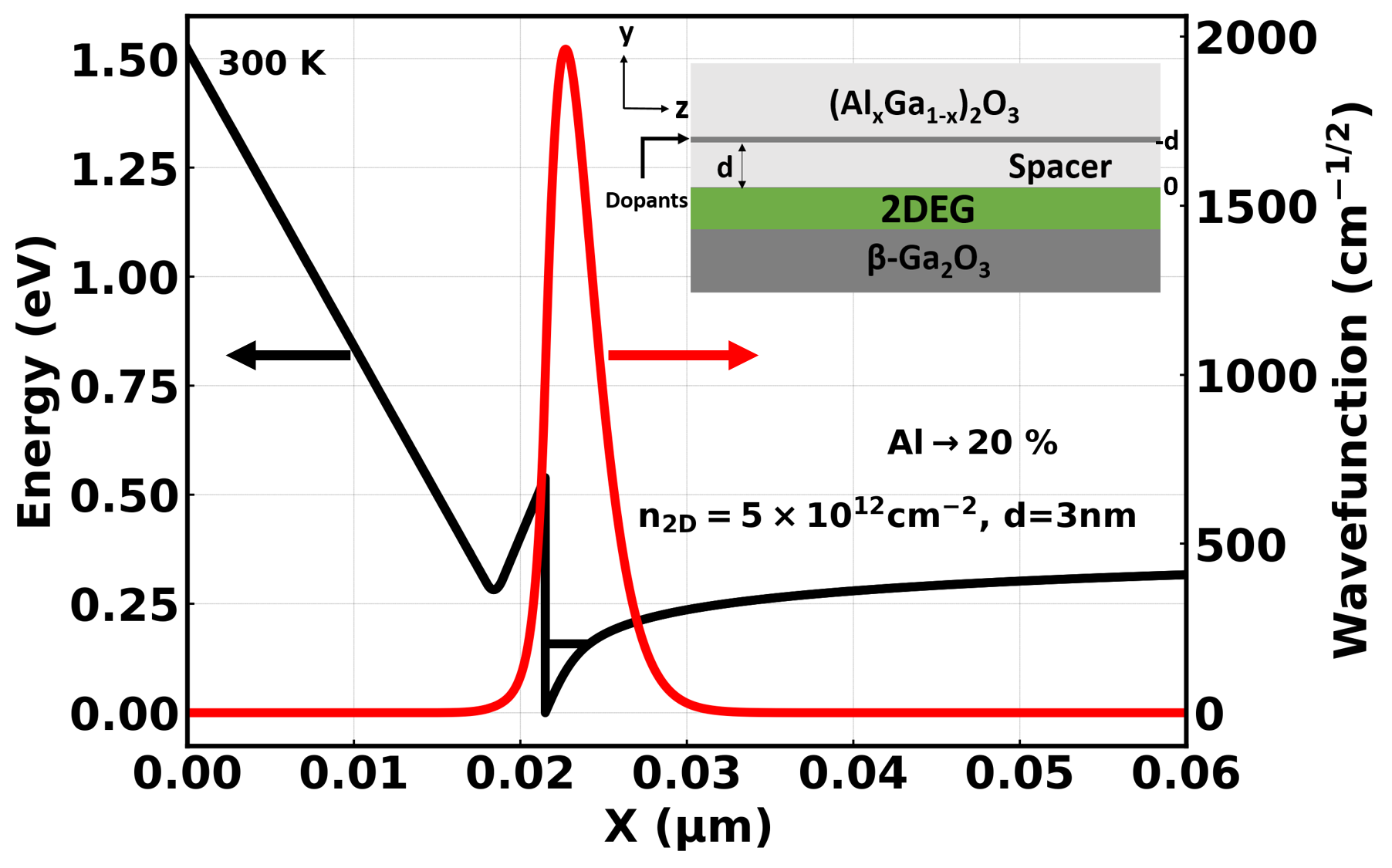}
\centering
\caption{Calculated energy band profile, first subband energy level and the corresponding wavefunction .The $\mathrm{\beta\mbox{-}(Al_{x}Ga_{1-x})_{2}O_{3}/Ga_{2}O_{3}}$ heterostructure is shown in the inset. The confinement direction is (010)}. \label{fig1}
\end{figure}

The 2DEG is realized at the interface of $\mathrm{\beta\mbox{-}(Al_{x}Ga_{1-x})_{2}O_{3}/Ga_{2}O_{3}}$  heterostructures as shown in the inset of fig.\ref{fig1}. The materials parameters used in this simulation are given in table \ref{tab:table1}. Here the region y$\geq$0 ($\mathrm{\beta\mbox{-}Ga_{2}O_{3}}$) is assumed to be unintentionally doped with the electron concentration of $\mathrm{1\times10^16}cm^{-3}$ cm$^{-3}$. The region -d$\leq$y$\leq$0 (spacer) is assumed to be free of any doping and the remote impurities are doped at y=-d ($\delta$ doping). Due to the difference in electron affinities of $\mathrm{\beta\mbox{-}(Al_{x}Ga_{1-x})_{2}O_{3}}$ and $\mathrm{\beta\mbox{-}Ga_{2}O_{3}}$, a 2DEG is formed at the interface. The confinement direction for the bench-line standard structure is cartesian y (crystal direction b of conventional unit cell) similar to the experimentally reported 2-DEGs grown on (010) Ga2O3 substrates\cite{RN18}. However, we also study the effect of confinement in other two cartesian directions (x and z) on low field electron mobility at room temperature.

 The electron gas confinement at the interface leads to the formation of energy subbands due to quantization of the electronic band structure. These quantized energy levels and the corresponding subbands are calculated by solving the Poisson's and Schrodinger's equations self-consistently. The total wavefunction and the energy eigenvalues of confined electronic states, characterized by subband index n and a 2D wave vector $\mathrm{\textbf{k}_{||}=(k_{x},k_{z})}$ parallel to the interface, are given as
\begin{equation}
    \mathrm{\Psi_{n,\textbf{k}}(\textbf{r},y)=\psi_{n}(y)exp(i\textbf{k}_{||}.\textbf{r})} 
\label{eqn:somelabel1}
\end{equation}
\begin{equation}
    \mathrm{E_{n}(\textbf{k})=E_{n}+\frac{\hbar^2k^2}{2m^*}}
    \label{eqn:somelabe2}
\end{equation}

where $\mathrm{\psi_{n}(y)}$ is the confined wavefunction, $\mathrm{m^*}$ is the electron effective mass in $\mathrm{\beta\mbox{-}Ga_{2}O_{3}}$, and $\mathrm{\hbar}$ is the reduced Planck's constant. $\mathrm{\psi_{n}(y)}$ satisfies the following Schrodinger's eq.,
\begin{equation}
    \mathrm{-\frac{\hbar^2}{2m^*}\frac{d^2\psi_{n}(y)}{dy^2}+V(y)\psi_{n}(y)=E_{n}\psi_{n}(y)}
\label{eqn:somelabel3}    
\end{equation}
where $\mathrm{V(y)=-e\phi_{e}(y)+V_{h}(y)}$ is the effective potential. Here $\mathrm{V_{h}(y)}$ is the step potential energy barrier at the interface, and $\mathrm{\phi_{e}(y)}$ is the electrostatic potential given by the solution of (Poisson's eq.) 
\begin{equation}
    \mathrm{\frac{d^2\phi_{e}(y)}{dy^2}=\frac{e}{\epsilon_{0}\epsilon}\Bigg[\sum_{i} N_{i}\psi_{i}^{2}(y)+N_{A}(y)-N_{D}(y)\Bigg]}
    \label{eqn:somelabel4}
\end{equation}
where,
\begin{equation}
    \mathrm{N_{i}=\frac{m^*k_{B}T}{\pi\hbar^2}ln\Bigg[1+exp\Big(\frac{E_{F}-E_{i}}{k_{B}T}\Big)\Bigg]}
    \label{eqn:somelabel5}
\end{equation}
Here $\mathrm{N_{i}}$ is the number of electrons in subband i. $\mathrm{N_{A}(y)}$ and $\mathrm{N_{D}(y)}$ are the acceptor and donor concentrations respectively. $\mathrm{E_{F}}$ is the fermi energy. We use Silvaco ATLAS to solve for the 2-DEG wavefunction \cite{RN30}.

The remote impurity doping density is adjusted in order to obtain a given electron density at different temperatures. The conduction band, 1st subband energy level and the corresponding wavefunction is shown in fig \ref{fig1}. From our calculations, we found that even at a 2DEG density of $\mathrm{5\times10^{12}cm^{-2}}$, the electrons only occupy the first subband, due to high density of states in $\mathrm{\beta\mbox{-}Ga_{2}O_{3}}$, at all assumed temperatures (10 K-500 K). Hence, we consider only the first subband in scattering rate calculations. We have not taken into account any phonon mediated intersubband scattering due to large separation between the first two sub-bands (70-80 meV) at assumed temperatures.

\begin{table}
\begin{center}
\begin{tabular}{ |c|c| } 
 \hline
 $\mathrm{m^*}$ & $\mathrm{0.3m_{e}}$  \\ 
  $\mathrm{E_{g} (Ga_{2}O_{3})}$ & $\mathrm{4.7}$ eV \cite{RN17}  \\ 
   $\mathrm{E_{g} ((Al_{0.2}Ga_{0.8})_{2}O_{3})}$ & $\mathrm{5.0}$ eV \cite{RN8} \\ 
    $\mathrm{\Delta E}$ & $\mathrm{0.54}$ eV \cite{RN25} \\ 
       $\mathrm{E_{d} ((Al_{0.2}Ga_{0.8})_{2}O_{3})}$ & $\mathrm{0.135}$ eV \cite{RN25}  \\ 
          $\mathrm{\epsilon_{r}}$ & $\mathrm{10}$ \cite{RN19,RN20}  \\ 
\hline
\end{tabular}
\caption{\label{tab:table1} Material Parameters used in the self-consistent calculation.}
\end{center}
\end{table}

\section{Scattering mechanisms}\label{scattering}
As discussed in section 1, the low field electron mobility in bulk $\mathrm{\beta\mbox{-}Ga_{2}O_{3}}$ is limited by polar optical phonons at room temperature. The same mechanism will play a significant role in the case of 2DEG too. In addition, we also include other important scatterings namely: remote impurity, non polar optical, acoustic deformation, alloy disorder, interface roughness and the residual impurity into our calculations. The 2-DEG wavefunction obtained from the self-consistent solution is used in the calculations of the intrasubband 2-D scattering rates. 

The screening of scattering potential becomes important in 2DEG when the electron density high. A careful treatment is required to include the screening in different scattering rates with good accuracy. Also, the effect of temperature on screening parameter needs to be taken into account while calculating the elastic and the non polar optical scattering rates. While, for polar optical modes the coupling with the 2-DEG plasmon and full dynamic screening is necessary for accurate calculations. We take into account the dynamic screening of LO phonons while a temperature dependent static screening is included for the other scattering mechanisms. 

\subsection{Static screening} \label{screening}
\begin{figure}[t]
\includegraphics[width=10cm]{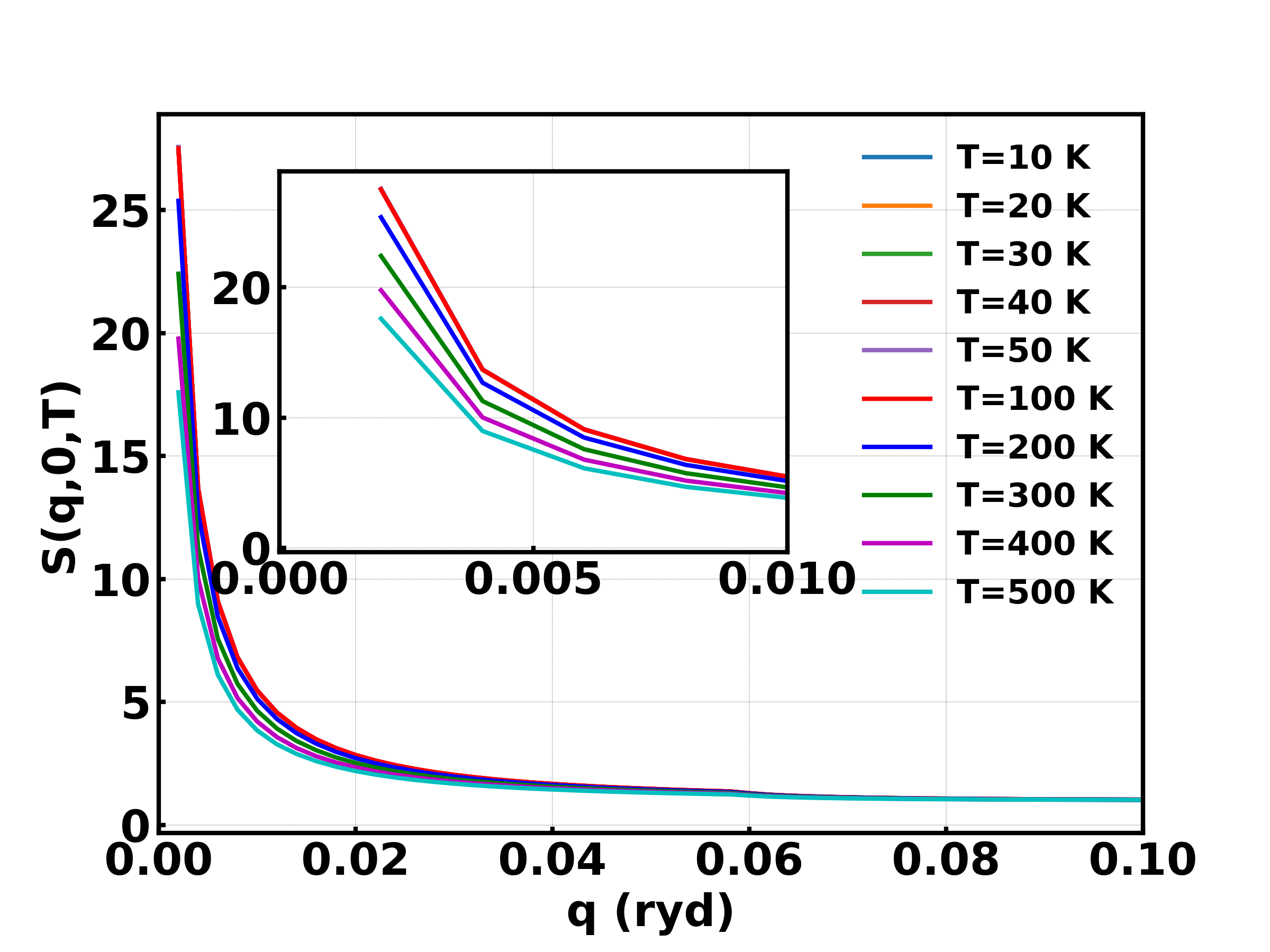}
\centering
\caption{Static screening for 2DEG at different temperatures. (Al=20$\%$, $\mathrm{n_{2D}=5\times10^{12}cm^{-2}}$, d=3 nm, $\mathrm{n_{RES}=1\times10^{16}cm^{-3}}$) }\label{fig2}
\end{figure}

In 2DEG, the remote impurity, alloy disorder and the interface roughness scatterings play a vital role in limiting the low field mobility at all temperatures. We expect these scattering potentials to get screened at very high electron density which strongly affects the low field electron mobility. Although a zero Kelvin static screening is widely used, we have included q-dependent and temperature dependent screening \cite{RN9}, given by 
\begin{equation}
    \mathrm{S(q) = 1+\frac{e^2F(q)\Pi(q)}{2\kappa_{0}\epsilon_{0}q}}\label{eqn:somelabel6}
\end{equation}
where the form factor, given by
\begin{equation}
    \mathrm{F(q) = \int_{0}^{\infty}dy\int_{0}^{\infty}dy^{\prime}|\psi_{n}(y)|^2|\psi_{n}(y^{\prime})|^2e^{-q|y-y^{\prime}|}dy}
    \label{eqn:somelabel7}
\end{equation}
takes into account the finite extent of the wavefunctions. The scattering potentials are divided by the obtained screening factor before the scattering strength (matrix elements) is calculated. Fig.\ref{fig2} shows the calculated screening factor at different temperatures as a function of q. It can be observed that screening is stronger for a small q and drops rapidly with increase in q.

\subsection{Phonon scattering}
We use the method developed by Ridley \cite{RN34} to calculate the interaction  ($\mathrm{\Big|M_{\bf{k^{\prime}k}}\Big|}$) between 2DEG and the bulk phonons. The calculated 2D matrix elements are then used to find the scattering rate under the Fermi's Golden rule \cite{RN23} as
\begin{equation}
    \mathrm{S_{\bf{kk^{\prime}}}=\sum_{\bf{k^{\prime}}} \frac{2\pi}{\hbar}\Big|M_{\bf{k^{\prime}k}}\Big|^2\delta(E(\bf{k^{\prime}})-E(\bf{k})\mp\hbar\omega)(1-f(\bf{k^{\prime}}))}
    \label{eqn:somelabel8}
\end{equation}
where $\mathrm{\Big|M_{\bf{k^{\prime}k}}\Big|}$ is the scattering strength, $\mathrm{E(\bf{k^{\prime}})}$, $\mathrm{E(\bf{k})}$ are the energy in the final and initial electronic state respectively. $\mathrm{\hbar\omega}$ is the phonon energy of a given mode. $\mathrm{(1-f(\bf{k^{\prime}}))}$ is the probability that the final state is empty.

 $\mathrm{\Big|M_{2D}(q_{||})\Big|^2=\sum_{q_{y}}\int_{-\infty}^{\infty}\Big|M_{3D}(q)\Big|^2|\psi_{n}(y)|^2e^{iq_{y}y}dy}$, is 2D scattering strength which takes care of the electron momentum conservation in the confinement direction. The momentum conservation in the 2DEG plane and the energy conservation is taken care by the numerical summation and the delta function in eq.\ref{eqn:somelabel7}. We assume a small energy smearing for 10 meV.

\subsubsection{Polar optical phonon scattering}
At high electron density, it is important to consider the dynamic screening of LO phonons .The interaction between the electric dipole moment and the electric field associated with the LO phonons and the plasmons respectively leads to couple modes with modified frequencies \cite{RN31,RN32,RN33}. The coupling is more pronounced when the plasmon and the phonon frerquencies have the same order. We take this coupling into account while calculating the screening contribution coming from the 12 IR active phonon modes and the high density plasmons.  

The vibrational energy of plasmon oscillations in a 2DEG is given by  
\begin{equation}
    \mathrm{\omega_{P}=\sqrt{\frac{\hbar^2n_{2D}e^2q}{m^*\epsilon_{\infty}}}}
    \label{eqn:somelabel9}
\end{equation}
where $\mathrm{n_{2D}}$ is the 2DEG density, q is the wavevector and $\mathrm{\epsilon_{\infty}}$ is the high frequency dielectric constant.

\begin{figure}[H]
\includegraphics[width=10cm]{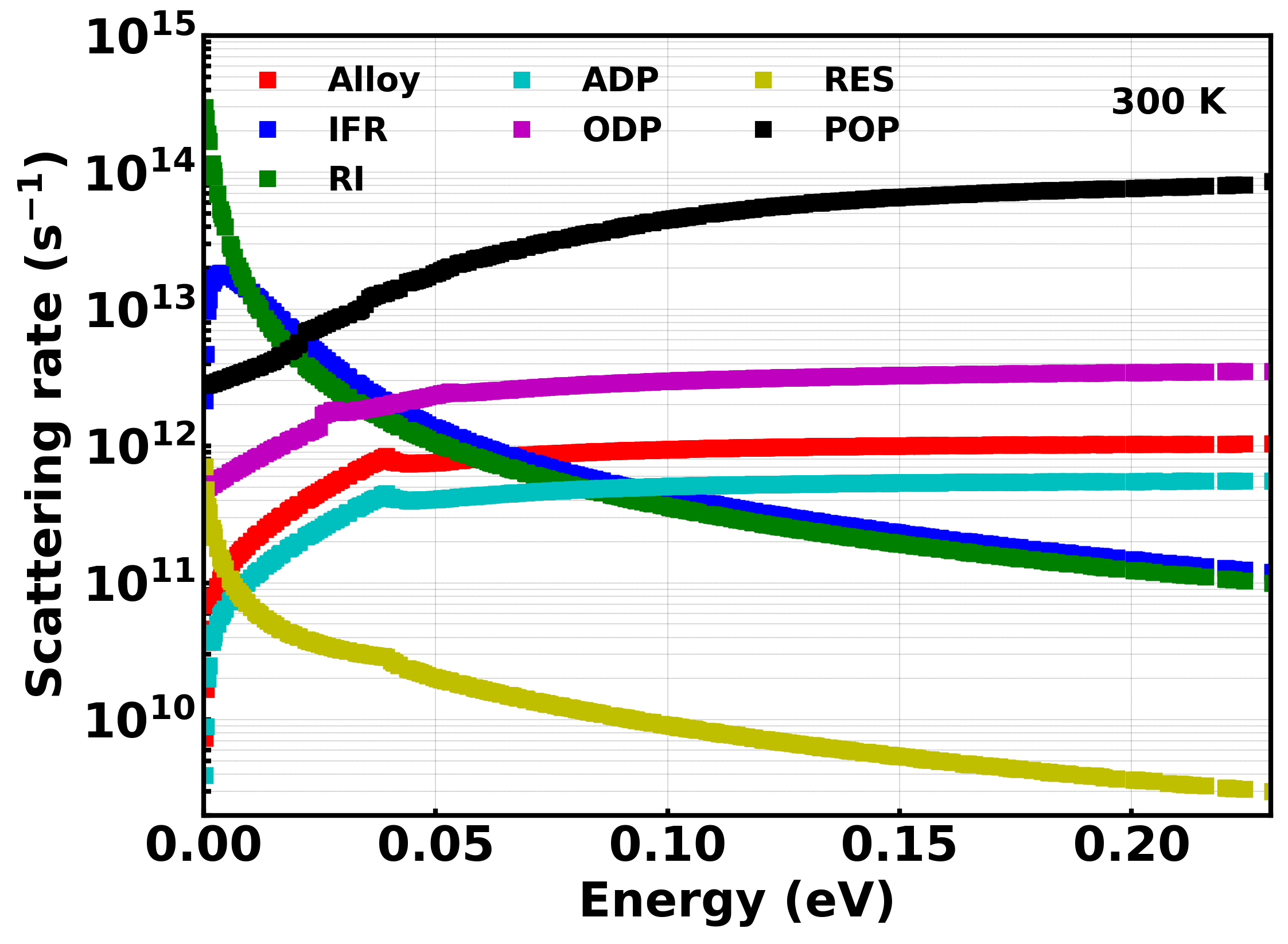}
\centering
\caption{Various scattering rates with electron energy. (Al = 20$\%$, $\mathrm{n_{2D}=5\times10^{12}cm^{-2}}$, d = 3 nm,    $\mathrm{n_{RES}=1\times10^{16}cm^{-3}}$, $\mathrm{\delta=0.5}$ nm, L= 5 nm) }\label{fig5}
\end{figure}

For high doping, the plasmon energy is in comparable range with the LO phonon energies and hence the LOPC plays a critical role in determining the free carrier screening. We calculate the LOPC modes corresponding to each q due to anisotropy of the LO modes. Following \cite{RN24}, we first compute the pure LO modes from LO-TO splitting for all q's and then the zeros of eq.\ref{eqn:somelabel10} \cite{RN1} gives the LOPC modes, corresponding to each q.
\begin{equation}
    \mathrm{\epsilon_{\omega}(\bf{q})=\epsilon_{\infty}\prod_{i=1}^{12}\frac{(\omega_{i}^{LO}(\bf{q}))^2-\omega^2}{(\omega_{i}^{TO})^2-\omega^2}-\frac{\epsilon_{\infty}\omega_{P}^2}{\omega^2}}\label{eqn:somelabel10}
\end{equation}
It has been observed that at low electron densities when the plasmon energy very low, the LOPC modes are of LO type. However, at higher electron densities,the high enough plasmon mode screens the splitting of LO-TO modes and such LOPC modes are mostly of TO type. Furthermore, we calculate the contribution of LO modes in each of the LOPC modes in order to calculate the scattering rate. Finally, to calculate the POP scattering strength from the Frohlich vertex for the electron-phonon interaction, we find a pair of dielectric constants $\mathrm{\varepsilon_{\omega_{v}^{LOPC}}^{+LOj}(q)}$ and $\mathrm{\varepsilon_{\omega_{v}^{LOPC}}^{-LOj}(q)}$ for each LO mode, for a given LOPC mode. The modified Frohlich eq. for the square of electron-phonon matrix elements is given as
\begin{equation}
    \mathrm{\Big|M_{LOPC}^{\nu,LOj}(\textbf{q})\Big|^2=\frac{e^2}{2\Omega\epsilon_{0}}\Bigg[\frac{\omega_{\nu}^{LOPC}(\textbf{q})}{q^2}\Bigg(\frac{1}{\epsilon_{\omega_{\nu}^{LOPC}}^{-LOj}(\textbf{q})}-\frac{1}{\epsilon_{\omega_{\nu}^{LOPC}}^{+LOj}(\textbf{q})}\Bigg)\Lambda_{\nu}^{LOj}(\textbf{q})\Bigg]}
    \label{eqn:somelabel11}
\end{equation}
Here $\mathrm{\Omega}$ is the unit cell volume, $\mathrm{\omega_{\nu}^{LOPC}(\textbf{q})}$ is the $\mathrm{\nu}$th LOPC frequency corresponding to a given q and $\mathrm{\Lambda_{\nu}^{LOj}(\textbf{q})}$ is the LO phonon content of the jth LO mode in the $\mathrm{\nu}$th LOPC mode. $\mathrm{\varepsilon_{\omega_{v}^{LOPC}}^{+LOj}(q)}$ includes the full response of that LO mode while $\mathrm{\varepsilon_{\omega_{v}^{LOPC}}^{-LOj}(q)}$ includes the response the all other modes keeping that LO mode frozen. Using the pair of dielectric constants, which take into account the LOPC screening , we first calculate the 3D matrix elements using eq.\ref{eqn:somelabel11}. The 2D scattering rate, as shown in fig.\ref{fig5}, is then calculated using eq. \ref{eqn:somelabel8}.

\subsubsection{Deformation potential scattering}
The acoustic and the non-polar scattering rates are usually ignored in the calculation of low field mobility for bulk $\mathrm{\beta\mbox{-}Ga_{2}O_{3}}$  as the polar optical scattering dominates. However, here in the case of 2DEG they become equally important and must be considered. We calculate these scattering rates based upon deformation potential approximation. To obtain the deformation potential (DP) constants we do a curve fitting with the 3D scattering rates calculated from the first principle at all assumed temperatures. Based on the ab-intio calculated density of states, for acoustic mode we obtain the following parameters $\mathrm{D_{A}^2/v_{s}^2 = 5\times10^{-11} eV^2s^2cm^{-2}}$, where $\mathrm{D_{A}}$ is the acoustic deformation potential and $\mathrm{v_{s}}$ is the velocity of sound. The non-polar optical deformation potential constants are given in table \ref{tab:table2}. The acoustic (ADP) and optical deformation potential (ODP) scattering rates are then calculated using 
\begin{equation}
    \mathrm{\frac{1}{\tau_{ADP}}=\frac{m^*k_{B}TD_{A}^2F_{nn}}{\pi\hbar^3v_{s}^2\rho}\int_{0}^{2\pi}\frac{1-cos\theta}{S(q)^2}d\theta}\label{eqn:somelabel12}
\end{equation}
\begin{equation}
    \mathrm{\frac{1}{\tau_{ODP}}=\frac{m^*D_{o}^2F_{nn}}{4\pi\hbar p\rho\omega_{o}}\Big(N_{\omega}+\frac{1}{2}\mp\frac{1}{2}\Big)\int_{q_{min}}^{q_{max}}\frac{dq}{S(q)^2Sin(\theta(q))}}
    \label{eqn:somelabel13}
\end{equation}
where p is the electron momentum, $\mathrm{\rho}$ is the mass density, $\mathrm{N_{\omega}}$ is the phonon occupancy, S(q) is the screening parameter, $\mathrm{cos(\theta(q))=\mp(\frac{\hbar q}{2p})+\frac{\omega}{vq}}$ and 
\begin{equation}
    \mathrm{F_{nn} = \int_{-\infty}^{\infty}|\psi_{n}(y)|^4dy}
    \label{eqn:somelabel14}
\end{equation}

From \cite{RN35,RN36}, we conclude that the short range deformation potential get screened the same way as long range phonon potential from free carriers. Since we are dealing with low field mobility, we assume a static screening (small q, degenerate electron gas), as described in section \ref{screening}. 

We then use the relaxation time approximation (RTA) to calculate the low field electron mobility for deformation potential and all other elastic scatterings \cite{RN23}.  

\begin{table}
\begin{center}
\begin{tabular}{ |c|c|c|c|} 
 \hline
 \bf{T(K)} & $\mathrm{\bf{D_{abs}^{2}/\hbar\omega_{o} (eVcm^{-2})}}$ & $\mathrm{\bf{D_{ems}^{2}/\hbar\omega_{o} (eVcm^{-2})}}$ & $\mathrm{\bf{\hbar\omega_{o} (eV)}}$ \\
 10 & $4.5\times10^{19}$ & $4.5\times10^{19}$ & 0.013 \\
  20 & $4.5\times10^{19}$ & $4.5\times10^{19}$ & 0.016 \\
   30 & $4.5\times10^{19}$ & $4.5\times10^{19}$ & 0.017 \\
    40 & $4.5\times10^{19}$ & $4.5\times10^{19}$ & 0.018 \\
     50 & $5.5\times10^{19}$ & $5.5\times10^{19}$ & 0.019 \\
      100 & $5.5\times10^{19}$ & $5.5\times10^{19}$ & 0.020 \\
       200 & $1.0\times10^{20}$ & $3.5\times10^{19}$ & 0.025 \\
        300 & $1.0\times10^{20}$ & $2.0\times10^{19}$ & 0.025 \\
         400 & $1.4\times10^{20}$ & $5.0\times10^{18}$ & 0.030 \\
          500 & $1.6\times10^{20}$ & $1.0\times10^{18}$ & 0.035 \\
 \hline
\end{tabular}
\caption{\label{tab:table2}Optical deformation potential constants (for ($\mathrm{\beta\mbox{-}Ga_{2}O_{3}}$)) calculated from first principle using fitting parameters at different temperatures.}
\end{center}
\end{table}


\subsection{Remote ionized impurity scattering}
The remote ionized impurities becomes  dominant at low temperature when the phonon occupancy is small in the crystal. The remote impurities are located in the $\mathrm{(Al_{x}Ga_{1-x})_{2}O_{3}}$ barrier layer. These are called remote impurities. The nature of the scattering rate due to these impurities depends on how far they are located. Although the impurity scattering rate is reduced, as compared to bulk, in case of 2DEG as they are located far from the electrons, they can have significant impact on the mobility depending upon the electron density and the spacer thickness. The remote impurity scattering rate is calculated using \cite{RN9}
\begin{equation}
    \mathrm{\frac{1}{\tau_{RI}}=\frac{m^*e^4Z^2}{8\pi\hbar^3\kappa_{0}^2\epsilon_{0}^2}\int_{0}^{2\pi}\Big(\frac{F(q,y_{i})}{qS(q)}\Big)^2N(y_{i})(1-cos\theta)dy_i}\label{eqn:somelabel15}
\end{equation}
where Ze is the charge on ionized impurities, $\mathrm{\kappa_{0}}$ is the static dielectric constant, $\mathrm{N(y_{i})}$ is the impurity distribution, and

\begin{equation}
    \mathrm{F(q,y_{i}) = \int_{-\infty}^{\infty}|\psi_{n}(y)|^2e^{-q|y_{i}-y|}dy}
    \label{eqn:somelabel16}
\end{equation}

\begin{figure}[H]
     \centering
     \begin{subfigure}[b]{0.49\textwidth}
         \centering
         \includegraphics[width=\textwidth]{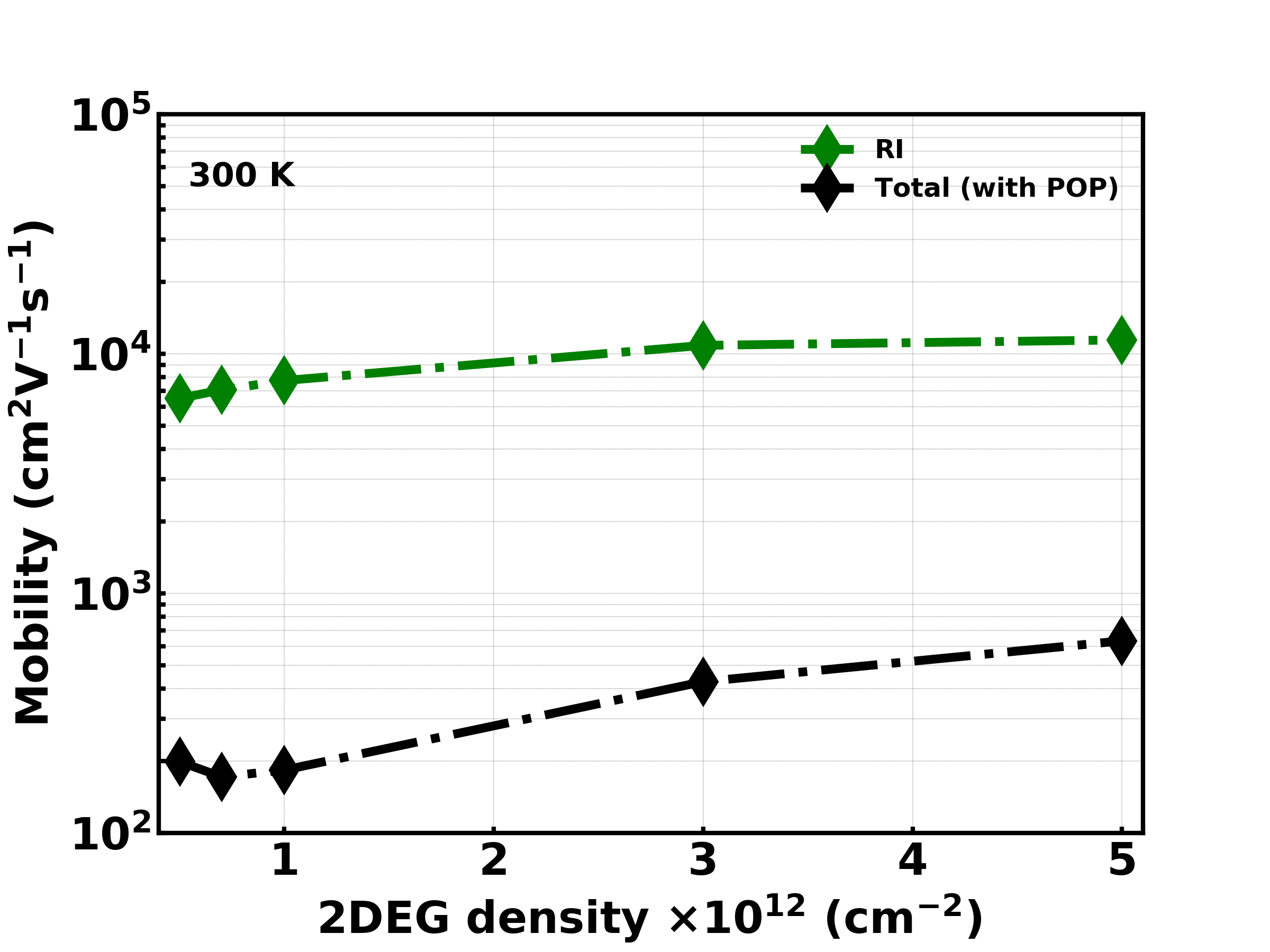}
         \caption{$\mathrm{\mu}$ Vs $\mathrm{n_{2D}}$.}
         \label{fig3a}
     \end{subfigure}
     \hfill
     \begin{subfigure}[b]{0.49\textwidth}
         \centering
         \includegraphics[width=\textwidth]{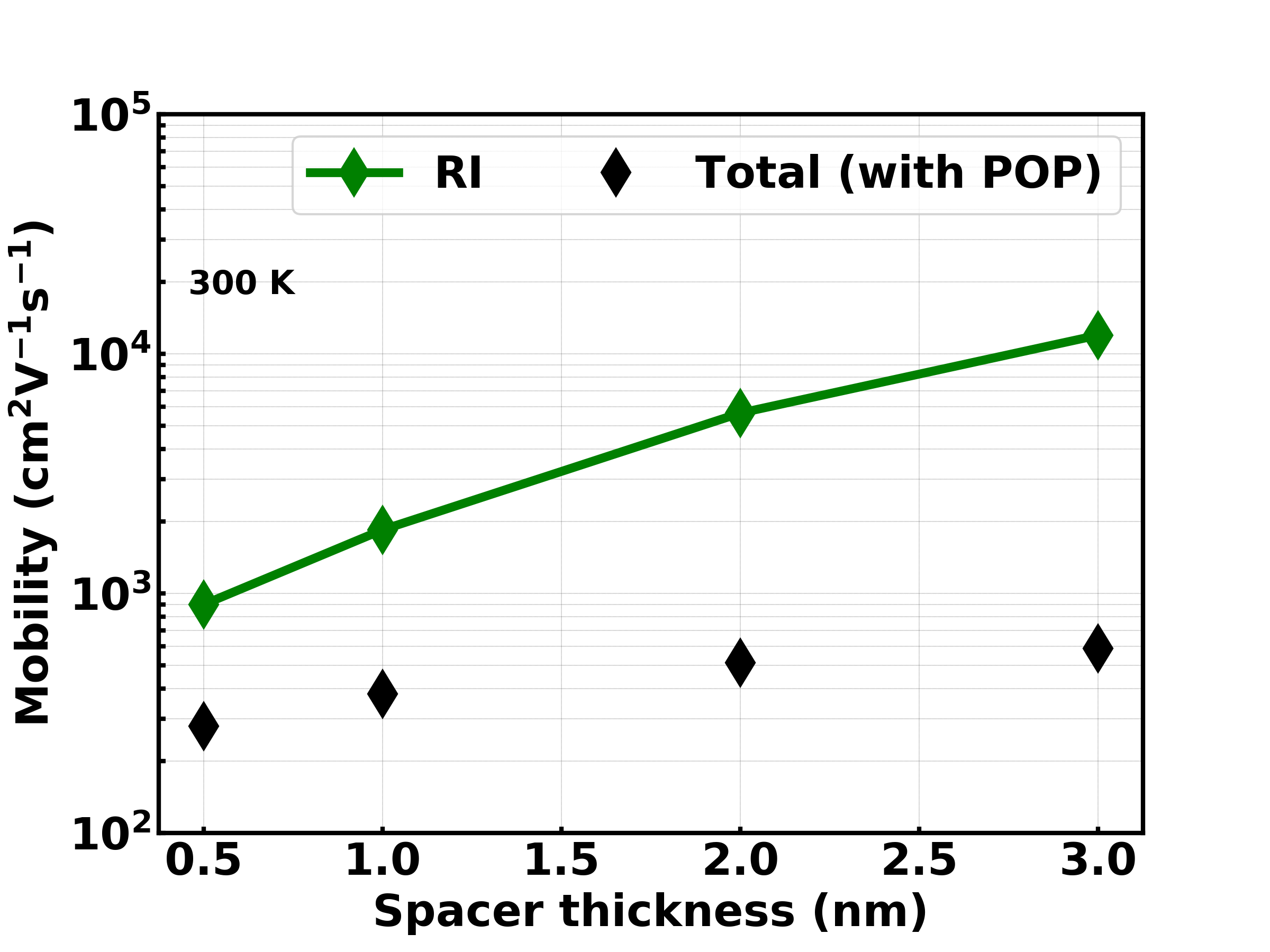}
         \caption{$\mathrm{\mu}$ Vs d.  }
         \label{fig3b}
     \end{subfigure}
        \caption{Low field electron mobility calculated using RTA as a function of 2DEG density (Al=20$\%$, d=3 nm, $\mathrm{n_{RES}=1\times10^{16}cm^{-3}}$, $\mathrm{\delta=0.5nm}$, L= 5nm) $\&$ spacer thickness (Al=20$\%$, $\mathrm{n_{2D}=5\times10^{12}cm^{-2}}$, $\mathrm{n_{RES}=1\times10^{16}cm^{-3}}$, $\mathrm{\delta=0.5nm}$, L= 5nm)}
        \label{fig3}
\end{figure}

At high enough density, screening becomes important. As shown in fig.\ref{fig3a}, the RI only impurity increases with 2DEG density which is due to better screening. We do not observe a significant improvement as with the increase in density, the form factor in eq.\ref{eqn:somelabel16}, which takes care of the interaction between the extent of wavefunction and the distributed impurity, becomes strong and hence weakens the screening. However, when the spacer thickness is small, the RI only mobility drops as the coulomb interaction becomes stronger as shown in fig.\ref{fig3b}. We observe a significant improvement in mobility (892-11967)$\mathrm{cm^2V^{-1}s^{-1}}$ when the thickness is increased from 0.5 nm to 3 nm. However, the larger thickness requires more impurity concentration which could lead to parallel channel formation as observed in \cite{RN22}. 

In addition to the remote impurities, we have assumed a low background doping of 1e16 cm-3 to calculate the residual impurity scattering. It becomes an important factor, as shown in experiments, depending upon the purity of the sample which in turn depends upon the growth and fabrication techniques. According to \cite{RN22,RN18}, the low temperature mobility on their sample is limited by high residual impurity.

\subsection{Interface-roughness scattering}
In heterostructures, it has been observed that the interface roughness plays an important role in limiting the mobility depending upon the characteristics of wavefunction in presence of electric field in the confinement direction. In theory, the roughness is statistically modeled using a autocovariance function which depends on two parameters called: the correlation length (L) and the rms height of the amplitude of the roughness ($\mathrm{\delta}$). Rougher interface means smaller L and larger delta. We calculate the interface roughness scattering rate using \cite{RN4}
\begin{equation}
    \mathrm{\frac{1}{\tau_{IFR}}=\frac{m^*e^2E_{eff}^2\delta^2 L^2}{2\hbar^3}\int_{0}^{2\pi}\frac{e^{-\frac{q^2L^2}{4}}(1-cos\theta)}{S(q)^2}d\theta}\label{eqn:somelabel17}
\end{equation}
where,

\begin{equation}
    \mathrm{E_{eff} = \int_{-\infty}^{\infty}|\psi_{n}(y)|^2\frac{dV}{dy}dy}
    \label{eqn:somelabel18}
\end{equation}
Here $\mathrm{\frac{dV}{dy}}$ is the Electric field along the confinement direction.

\begin{figure}[H]
     \centering
     \begin{subfigure}[b]{0.49\textwidth}
         \centering
         \includegraphics[width=\textwidth]{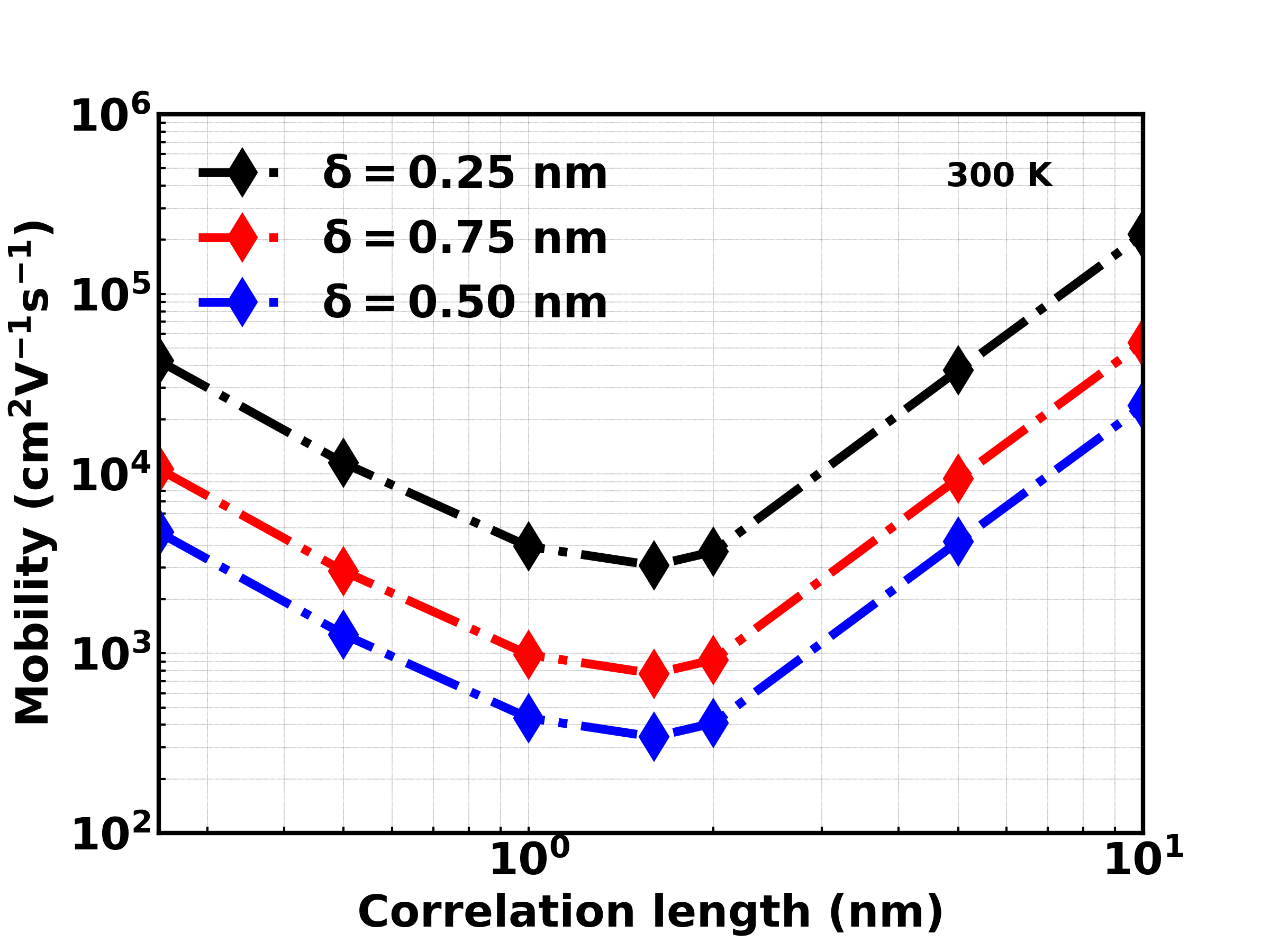}
         \caption{$\mathrm{\mu}$ Vs L.}
         \label{fig4a}
     \end{subfigure}
     \hfill
     \begin{subfigure}[b]{0.49\textwidth}
         \centering
         \includegraphics[width=\textwidth]{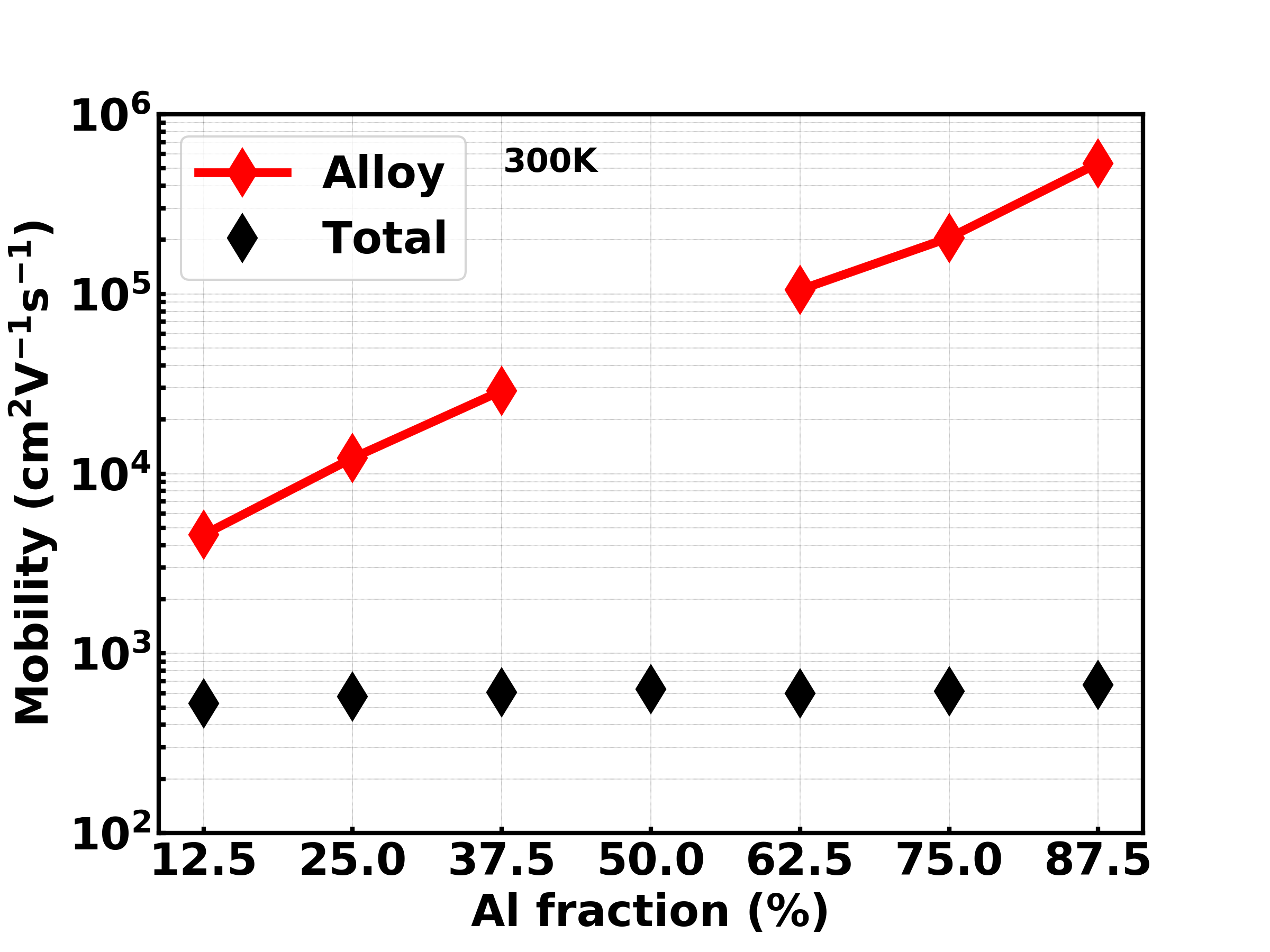}
         \caption{$\mathrm{\mu}$ Vs Al$\%$.}
         \label{fig4b}
     \end{subfigure}
        \caption{Low field electron mobility calculated using RTA as a function of correlation length (Al=20$\%$, $\mathrm{n_{2D}=5\times10^{12}cm^{-2}}$, d=3 nm, $\mathrm{n_{RES}=1\times10^{16}cm^{-3}}$) $\&$ Al fraction ($\mathrm{n_{2D}=5\times10^{12}cm^{-2}}$, d=3 nm, $\mathrm{n_{RES}=1\times10^{16}cm^{-3}}$, $\mathrm{\delta=0.5nm}$, L= 5nm)}
        \label{fig4}
\end{figure}

The IFR only mobility (300K) as a function of correlation length (L), at three different $\mathrm{\delta}$, is shown in fig.\ref{fig4a}. We observe a minimum mobility when the correlation length ($\mathrm{L_{min}}$) ($\mathrm{\lambda_{F}}$) is equal to the fermi wavelength for each $\mathrm{\delta}$. When L is decreased from $\mathrm{L_{min}}$, the electrons with higher fermi wavelength cannot see the lateral roughness and hence the mobility keeps increasing whereas when L is increased from $\mathrm{\lambda_{F}}$, the roughness becomes ineffective (large L) again and hence the mobility keeps increasing. We use typical roughness parameters ($\mathrm{\delta}$)=0.5 nm and L= 5 nm for the rest of our standard calculations. 

\subsection{Alloy disorder scattering}
It is obvious from quantum mechanics that due to a finite potential barrier between the $\mathrm{(Al_{x}Ga_{1-x})_{2}O_{3}}$ and $\mathrm{(Ga_{2}O_{3}}$ (depending upon there electron affinities), the wavefunction will penetrate across the barrier and interact with the disorder in alloy. We calculate the alloy disorder scattering rate using eq.\ref{eqn:somelabel19}. The volume $\mathrm{\Omega_{o}}$ of the unit cell for $\mathrm{(Al_{x}Ga_{1-x})_{2}O_{3}}$ for different alloy concentration is taken from \cite{RN8}. The scattering potential is assumed to be equal to the conduction band offset $\mathrm{\delta E_{c}}$ between $\mathrm{(Al_{2}O_{3}}$ and $\mathrm{(Ga_{2}O_{3}}$ (2.67 eV) \cite{RN8}.

\begin{equation}
    \mathrm{\frac{1}{\tau_{Alloy}}=\frac{m^*\Omega_{o}x(1-x)(\delta E_{c})^2F_{al}}{2\pi\hbar^3}\int_{0}^{2\pi}\frac{1-cos\theta}{S(q)^2}d\theta}\label{eqn:somelabel19}
\end{equation}

\begin{equation}
    \mathrm{F_{al} = \int_{-\infty}^{0}|\psi_{n}(y)|^4dy}
    \label{eqn:somelabel20}
\end{equation}

We show, in fig.\ref{fig4b}, the Alloy only mobility with the Aluminium $\%$ in the alloy. We see an improvement in the mobility as we go from (12.5-87.5)$\%$ Al. This is because higher fraction of Al increases the barrier height at the interface which reduces the wavefunction leakage into the alloy and hence the scattering rate (form factor). At 50$\%$ Al concentration, the alloy is ordered and hence we don't observe any alloy disorder scattering. The low field mobility is observed to be lower than in GaN. This is because of larger unit cell volume and higher scattering potential ($\mathrm{\delta E_{c}}$) \cite{RN38}.

\section{Low field electron transport}\label{mobility}
The scattering rates described above are calculated on a 2D k-space and then we solve the boltzmann transport equation using Rode's iterative method. We assume the non-polar optical scattering rate to be isotropic to be able to use Mathiessen's sum rule. The convergence in the perturbation in distribution function, given by \ref{eqn:somelabel21}, is achieved in few iterations \cite{RN13}. 
\begin{equation}
    \mathrm{g_{k,i+1}=\Bigg(\frac{S_{i}(g_{k},i)-\frac{eF}{\hbar}\frac{\partial f}{\partial k}}{S_{o}(k)+\frac{1}{\tau_{el}(k)}}}\Bigg)
    \label{eqn:somelabel21}
\end{equation}
where $\mathrm{S_{i}}$ and $\mathrm{S_{o}}$ are the in-scattering and out-scattering rates respectively \cite{RN13}, $\mathrm{\frac{1}{\tau_{el}}}$ is the total elastic scattering rates and F is the applied electric field in a given direction.

\subsection{Mobility}
We calculate an average low field electron mobility mobility using the method described in \cite{RN2}, using eq.(\ref{eqn:somelabel22}) 
\begin{equation}
    \mathrm{\mu_{n}=\frac{m^*}{2\pi\hbar^2Fn_{s}}\int_{0}^{\infty}g(E)v(E)dE}\label{eq:mu}
    \label{eqn:somelabel22}
\end{equation}
Here v(E) is the velocity of electrons calculated from spherical band approximation.

\section{Results and Discussion}\label{result}
We expect the deformation potential and the alloy disorder scattering to stay constant with the increase in electron energy as they are isotropic in nature and follow the variation in density of states. However, as shown in fig\ref{fig5}, we see an increase in the scattering rate which is coming from the improvement in the screening with decrease in q. The strong inverse dependence on q makes the remote and the residual impurity scattering anisotropic and so they are decreasing with increase in electorn energy. We see a peak at very small energy in IFR scattering rate due to the interplay between the inverse-exponential (autocorrelation function) dependence and the screening. We observe a usual trend in the variation of POP scattering rate.

\begin{figure} [H]
     \centering
     \begin{subfigure}[b]{0.49\textwidth}
         \centering
         \includegraphics[width=\textwidth]{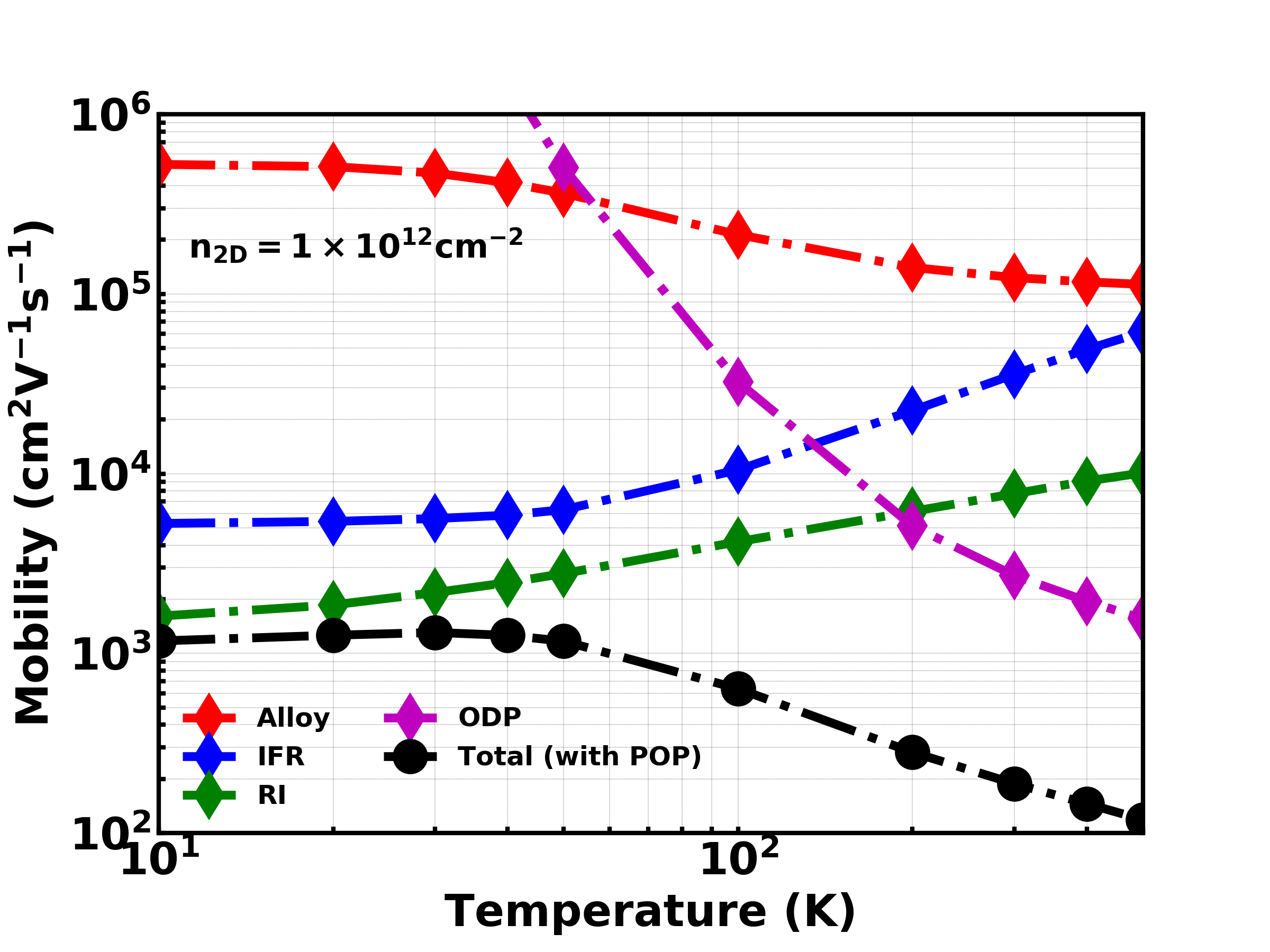}
         \caption{$\mathrm{\mu}$ Vs T. $\mathrm{n_{2D}=1\times10^{12}cm^{-2}}$}
         \label{fig6a}
     \end{subfigure}
     \hfill
     \begin{subfigure}[b]{0.49\textwidth}
         \centering
         \includegraphics[width=\textwidth]{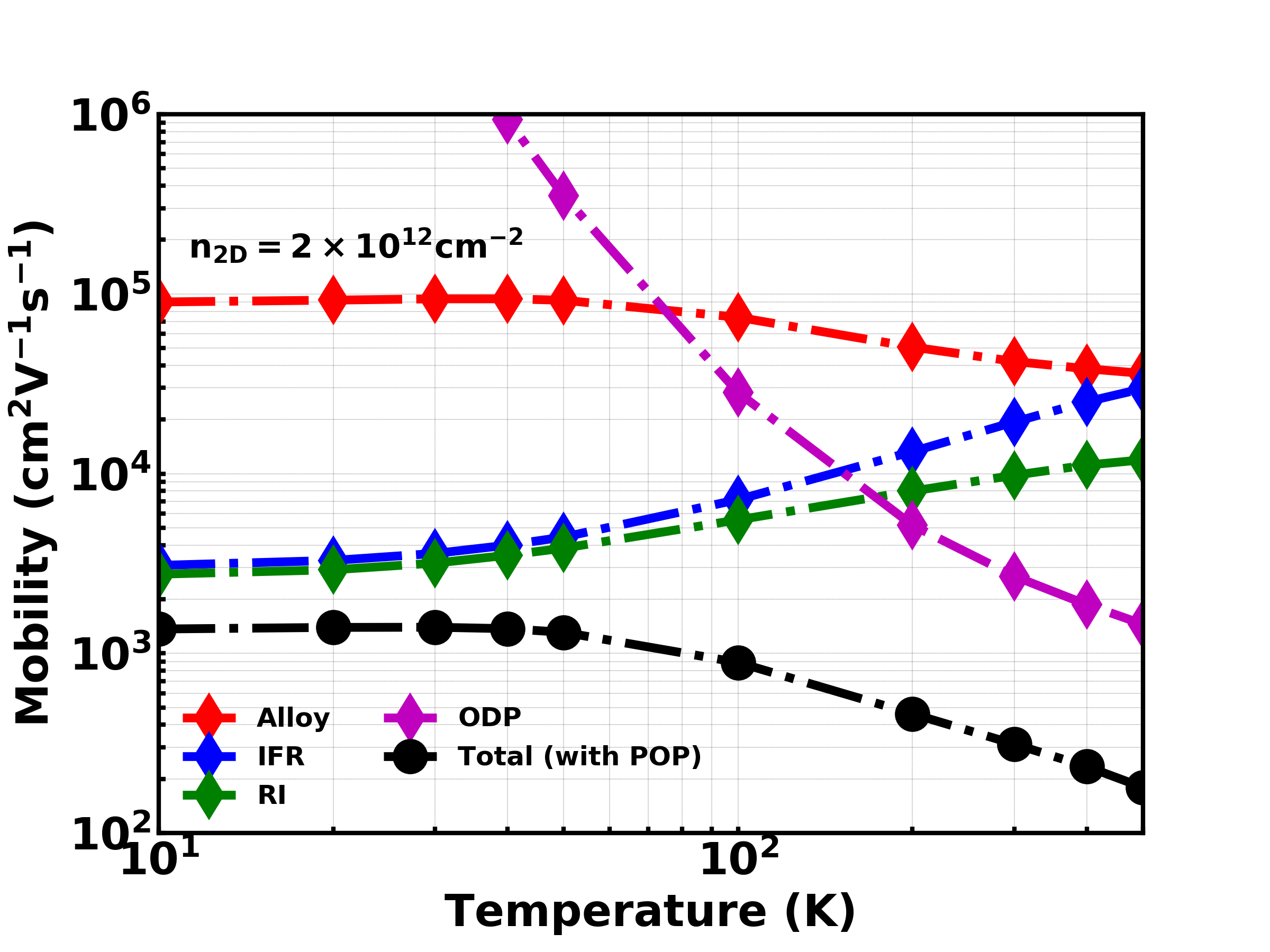}
         \caption{$\mathrm{\mu}$ Vs T. $\mathrm{n_{2D}=2\times10^{12}cm^{-2}}$}
         \label{fig6b}
     \end{subfigure}
     \hfill
     \begin{subfigure}[b]{0.49\textwidth}
         \centering
         \includegraphics[width=\textwidth]{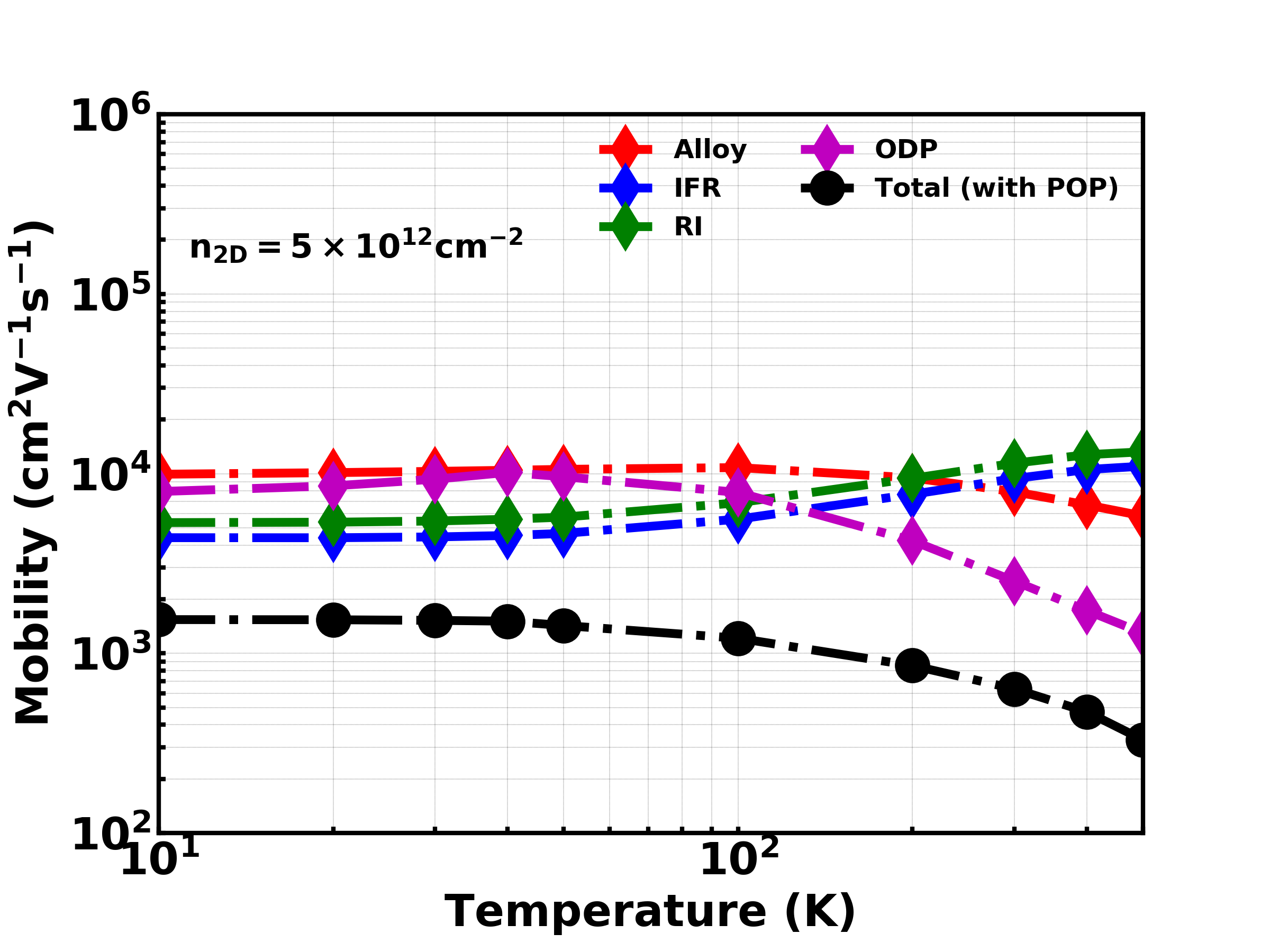}
         \caption{$\mathrm{\mu}$ Vs T. $\mathrm{n_{2D}=5\times10^{12}cm^{-2}}$}
         \label{fig6c}
     \end{subfigure}
        \caption{Low field electron mobility calculated using Rode's iterative method as a function of temperature. (Al=20$\%$, d=3 nm, $\mathrm{n_{RES}=1\times10^{16}cm^{-3}}$, $\mathrm{\delta=0.5nm)}$, L= 5nm }
        \label{fig6}
\end{figure}

The temperature dependence of low field electron mobility, at three different electron densities, is shown in fig\ref{fig6}. The trend in the variation of mobility with temperature can be explained from the variation in scattering rates with electron energy and given the fact that the electrons will dispersed to higher energy with the increase in temperature. The total mobility in the high temperature regime is dominated by the polar optical phonon scattering but at the same time seem to be enhanced when compare to bulk, thanks to the dynamic screening of LO phonons. The interplay between the screening factor and the form factor determines the dependence of mobility on 2DEG density. The screening decreases however the form factor increases the scattering rate. We observe that the contribution to the mobility from the remote impurities shifts upwards with the increase in electron density due to better screening. The alloy disorder mobility shifts downwards as more electron density leads to more leakage into the alloy. The interplay between the increase in the effective electric field at the interface and the screening, with the increase in electron density, leads the IFR mobility to first shift downwards and then upwards. The optical deformation potential mobility decreases rapidly at low temperature, with increase in density, as the fermi level is high enough so that the majority of electrons can both absorb and emit a given phonon. The remote impurity and the surface roughness scatterings seem to be highly dominating in the low temperature regime. The residual impurity doesn't seem to have a much impact on the mobility. 

\begin{figure}[H]
\includegraphics[width=10cm]{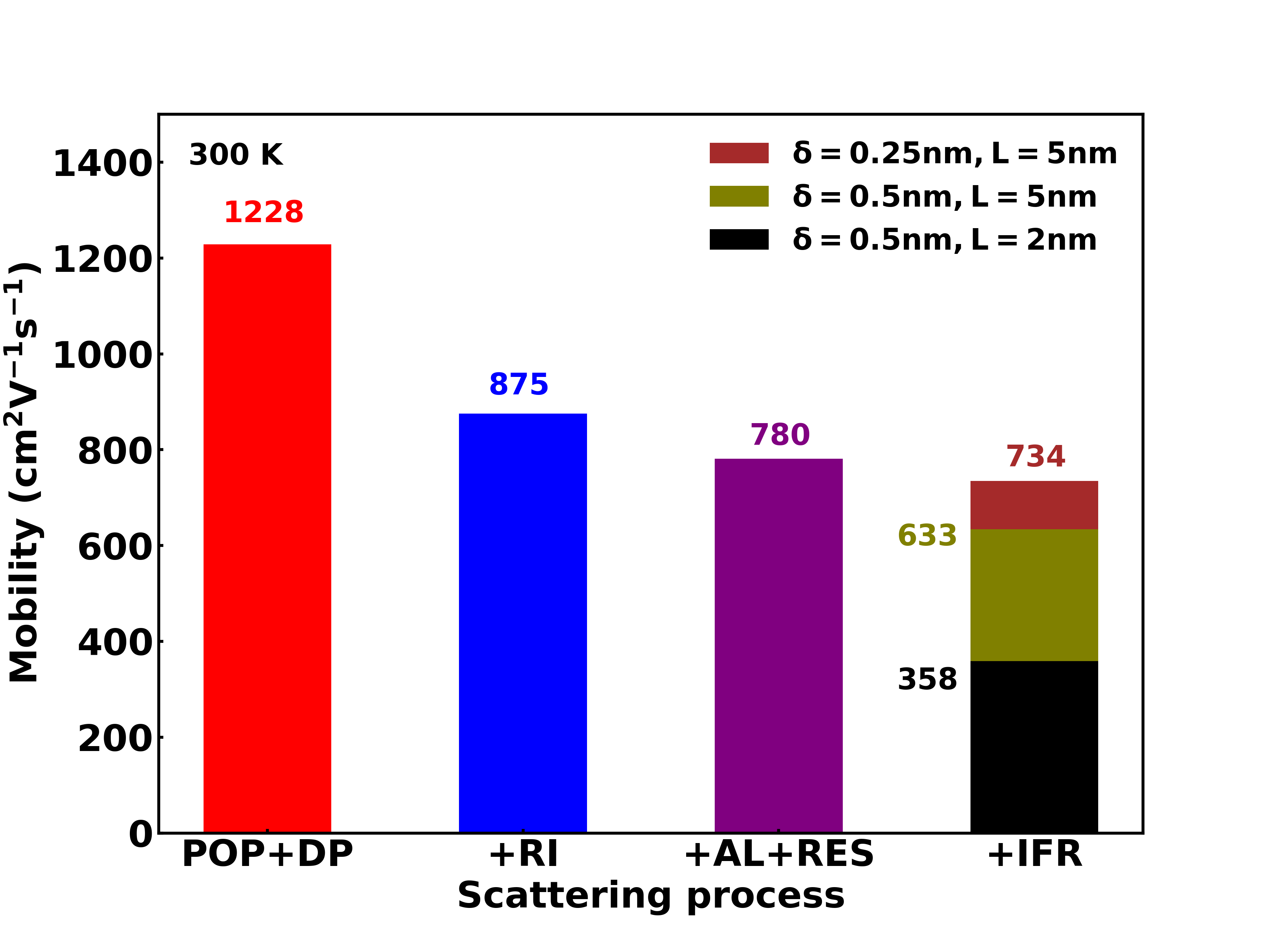}
\centering
\caption{Total low field electron mobility as a function of different scattering mechanisms. (Al=20$\%$, $\mathrm{n_{2D}=5\times10^{12}cm^{-2}}$ , d=3 nm, $\mathrm{n_{RES}=1\times10^{16}cm^{-3}}$)}\label{fig7}
\end{figure}

In fig.\ref{fig7}, we show the effect on mobility with the addition of different scattering processes. The total mobility is found to be 1228 $\mathrm{cm^2V^{-1}s^{1}}$ when only intrinsic phonon contribution is considered. The mobility drops to 780 $\mathrm{cm^2V^{-1}s^{1}}$  with the addition of Remote, alloy disorder, residual and IFR scattering processes. We then show the total mobility including three different levels of roughness. The total mobility with roughness L=5 nm $\&$ $\delta$=0.5 nm, when the optical deformation potential screening is not included, is found to be 523 $\mathrm{cm^2v^{-1}s^{-1}}$. 

The total mobility as a function of 2DEG density is shown in fig.\ref{fig3a}. The overall mobility seems to be increasing due to increase in screening. We also observe an anti-screening when the density is increases from $\mathrm{5\times10^{11}cm^{-2}}$ to $\mathrm{7\times10^{11}cm^{-2}}$ coming from the coupled phonon-plasmon modes. There is also an increase in the total mobility in fig.\ref{fig3b} when the spacer thickness is increases which is because the remote impurity becomes less effective. We observe almost no change in the total mobility, as shown in fig.\ref{fig4b}, when the Al fraction is increased as at higher alloy percentage, the POP and ODP scatterings are dominant at 300K.  

\begin{figure}[H]
\includegraphics[width=10cm]{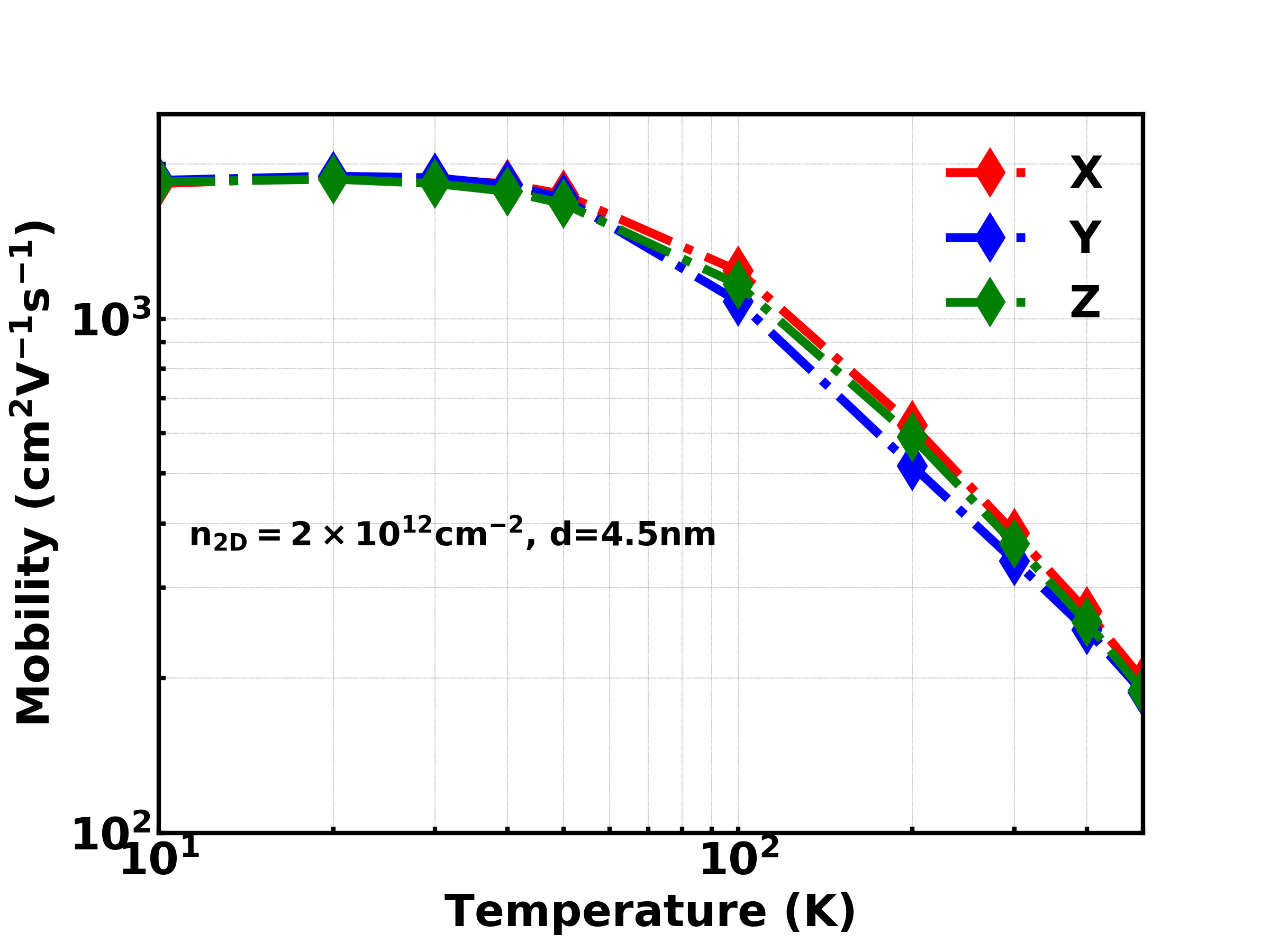}
\centering
\caption{$\mathrm{\mu}$ Vs T for different confinement direction. (Al=20$\%$, $\mathrm{n_{RES}=1\times10^{16}cm^{-3}}$, $\mathrm{\delta=0.5}$ nm, L= 5 nm )}\label{fig8}
\end{figure}

The variation in total mobility with temperature for different confinement directions is shown in fig\ref{fig8}. In order to clearly see the effect of confinement because of POP only scattering, we reduce the 2DEG density to $\mathrm{1\times10^{12}cm^{-2}}$ and hence the dynamic screening and increase the spacer thickness to 4.5 nm. Depending on the dominant mode (Au or Bu) at a given temperature, the mobility is higher or lower. Since, as shown in \cite{RN5}, Bu modes (polarized in cartesian x and z directions) are dominant at room temperature, we see a very little improvement in the mobility when the 2DEG is confined in either cartesian x or z.


\section{Summary}\label{summary}
The low field electron mobility in the 2DEG of $\mathrm{\beta\mbox{-}(Al_{x}Ga_{1-x})_{2}O_{3}/Ga_{2}O_{3}}$ heterostructures is calculated and found to be improved compared to bulk. The low temperature mobility is found to be dominated by interface roughness and alloy disorder scattering. Due to very high 2DEG density, the ODP scattering also plays a significant role. At 300 K, the total mobility is dominated by POP $\&$ ODP scattering and decreases significantly based on roughness parameters.  

\section{Acknowledgement}
The authors acknowledge the support from Air Force Office of Scientific Research under award number FA9550-18-1-0479 (Program Manager: Ali Sayir) and from NSF under awards ECCS 1607833 and ECCS - 1809077. The authors also acknowledge the high performance computing facility provided by the Center for Computational Research (CCR) at University at Buffalo.  

\bibliography{References}
\bibliographystyle{unsrt}

\end{document}